\documentclass[sigconf]{acmart}

\usepackage{etoolbox}
\patchcmd{\quote}{\rightmargin}{\leftmargin 2em \rightmargin}{}{}

\usepackage{balance}
\usepackage{enumitem}
\usepackage{url}
\usepackage{color}
\usepackage{tikz}
\usetikzlibrary{arrows, positioning, shapes.geometric, }

\tikzset{flowchart/.style = {
     base/.style = {draw=black, 
                    inner sep=1mm, outer sep=0mm,
                    text width=3cm, minimum height=0.5cm,
                    align=flush center},
  process/.style = {base, fill=green!30},
       io/.style = {base, trapezium,
                    trapezium left angle=70, trapezium right angle=110,
                    trapezium stretches=true, 
                    text width=1.5cm,
                    fill=orange!30},
     arrow/.style = {thick,->,>=stealth}
                    }}

\newcommand{\ignore}[1]{}

\copyrightyear{2017}
\acmYear{2017}
\setcopyright{rightsretained}
\acmConference{SIGIR 2017 Workshop on Neural Information Retrieval (Neu-IR'17)}{}{\\August 7--11, 2017, Shinjuku, Tokyo, Japan} 
\acmPrice{}
\acmDOI{}
\acmISBN{}
\acmPrice{}

\begin{document}

\title[]{Exploring the Effectiveness of Convolutional Neural Networks\\ for Answer Selection in End-to-End Question Answering}

\author{Royal Sequiera,$^1$ Gaurav Baruah,$^1$ Zhucheng Tu,$^1$ Salman Mohammed,$^1$ \\Jinfeng Rao,$^2$ Haotian Zhang,$^1$ and Jimmy Lin$^1$}
\affiliation{\vspace{0.1cm}
\institution{$^1$ David R. Cheriton School of Computer Science, University of Waterloo}
\institution{$^2$ Department of Computer Science, University of Maryland}
}
\email{{rdsequie,gbaruah,michael.tu,salman.mohammed,haotian.zhang,jimmylin}@uwaterloo.ca, jinfeng@cs.umd.edu}

\begin{abstract}
Most work on natural language question answering today focuses on
answer selection:\ given a candidate list of sentences, determine
which contains the answer. Although important, answer selection is
only one stage in a standard end-to-end question answering
pipeline. This paper explores the effectiveness of convolutional neural
networks (CNNs) for answer selection in an end-to-end context using
the standard TrecQA dataset. 
We observe that a simple {\it idf}-weighted word overlap
algorithm forms a very strong baseline, and that despite substantial
efforts by the community in applying deep learning to tackle answer selection,
the gains are modest at best on this dataset.
Furthermore, it is unclear if a CNN is more effective than the baseline
in an end-to-end context based on
standard retrieval metrics. To further explore this finding, we conducted
a manual user evaluation, which confirms that
answers from the CNN are detectably better than those from {\it idf}-weighted
word overlap. This result suggests that users are sensitive to relatively small
differences in answer selection quality.
\end{abstract}

\settopmatter{printacmref=false, printfolios=false}

\maketitle

\section{Introduction}

Natural language question answering (QA) over free text has a long
history that dates back many decades, but most recent
studies---especially those based on deep learning---focus almost exclusively
on the answer selection problem, which is one stage in an
end-to-end pipeline. Given a question and a number of candidate
sentences, the answer selection task is to decide which of the
sentences contains the correct answer. Of course, these candidates
have to come from {\it somewhere} and {\it somehow}.

Quite naturally, candidate sentences for answer selection originate
from a document collection, and are typically identified based on
document retrieval and some term-based passage extraction scheme. Yet,
these important parts of the QA pipeline are not considered in most
modern evaluations---most QA datasets today encapsulate only answer
selection.

In this paper, we examine the effectiveness of answer selection as a
component in an end-to-end question answering system, using the
widely-used TrecQA dataset. The contribution of this work lies in
three interesting findings:

\begin{itemize}[leftmargin=*]

\item Experiments on the TrecQA dataset show that scoring sentences based on {\it
  idf}-weighted word overlap forms a very strong baseline,
  and that the gap between this baseline and the state of the art is
  surprisingly small. This is not a new finding, although this result
  does not appear to be widely known in the literature. Despite
  substantial effort, mostly by the natural language processing
  community, the gains from deep learning are modest at best.

\item When examining the effectiveness of a standard convolutional
  neural network for answer selection in an end-to-end context, it is
  not clear if the neural network is better than the {\it idf}-weighted word overlap
  baseline according to standard IR evaluation metrics. This can be interpreted as a negative result.

\item To further explore the previous finding, we conducted a manual
  evaluation, which showed that the output of the convolutional neural
  network is indeed detectably better (by humans) than the simple {\it idf}
  baseline. This suggests that end users are quite sensitive
  to relatively small differences in answer selection quality.

\end{itemize} 

\noindent Taken together, these findings show the importance of
conducting both component-level evaluations (answer selection) as well as
end-to-end evaluations. The latter is ignored in most studies today,
which we feel is a major oversight. We recommend that moving forward, such
end-to-end evaluations be given more prominence.

\section{Background and Related Work}

\subsection{Question Answering Architectures}

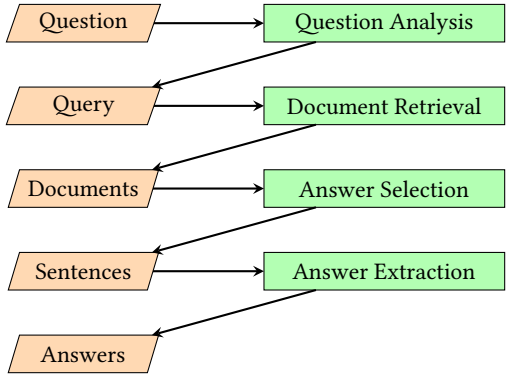
\begin{figure}
\vspace{0.7cm}
\begin{tikzpicture}[flowchart,
    node distance=2cm, auto
                    ]
\node (question) [io] {Question} ;
\node (analyzer) [process, right of=question, xshift=2cm] {Question Analysis};
\node (query) [io, below of=question, yshift=+0.9cm] {Query};
\node (docRetriever) [process, right of=query, xshift=2cm] {Document Retrieval};
\node (documents) [io, below of=query, yshift=+0.9cm] {Documents};
\node (passageRetriever) [process, right of=documents, xshift=2cm] {Answer Selection};
\node (passages) [io, below of=documents, yshift=+0.9cm] {Sentences};
\node (answerExtractor) [process, right of=passages, xshift=2cm] {Answer Extraction};
\node (answers) [io, below of=passages, yshift=+0.9cm] {Answers};

\draw [arrow] (question) -- (analyzer);
\draw [arrow] (analyzer) -- (query);
\draw [arrow] (query) -- (docRetriever);
\draw [arrow] (docRetriever) -- (documents);
\draw [arrow] (documents) -- (passageRetriever);
\draw [arrow] (passageRetriever) -- (passages);
\draw [arrow] (passages) -- (answerExtractor);
\draw [arrow] (answerExtractor) -- (answers);
\end{tikzpicture}
\caption{A typical question answering pipeline architecture, adapted
  from Tellex et al.~\cite{Tellex_etal_SIGIR2003}.}
\label{fig:e2e-pipeline}
\end{figure}

Given a question $q$ and a candidate set of sentences $\{c_1, c_2,
\ldots c_n\}$, the answer selection task is to identify sentences that
contain the answer. Answer selection forms an important
component in the standard pipeline architecture for question answering depicted in
Figure~\ref{fig:e2e-pipeline}, adapted from Tellex et
al.~\cite{Tellex_etal_SIGIR2003}. Although details vary from system to
system, a general QA architecture consists of a question analysis component to
convert the natural language question into a search query, a document
retrieval component to fetch a set of documents, and an answer selection component to
identify the best sentences (or more generally, passages). In
some designs, an answer extraction component identifies the exact
natural language phrase that answers the question~\cite{Voorhees_TREC2002,Lin_etal_CHI2003}.

In this setup, answer selection putatively works on candidate sentences
retrieved from the document collection.
Although nominally a classification task, answer selection is usually
evaluated in terms of ranked retrieval metrics. In other words, answer
selection can be viewed as reranking the output of sentences from a previous stage in the pipeline,
similar to multi-stage ranking architectures in the web
context~\cite{Pedersen_SIGIR2010,Wang_etal_SIGIR2011,Tonellotto_etal_WSDM2013,Asadi_Lin_SIGIR2013,Culpepper_etal_ADCS2016}.
The literature also refers to this as a ``telescoping'' setup~\cite{Matveeva_etal_SIGIR2006}, which has emerged as
a standard way to evaluate neural ranking models~\cite{Mitra_Craswell_2017}. Thus, although
our work examines only question answering, our findings are likely applicable
to a broad range of information retrieval tasks.

\subsection{CNN for Answer Selection}

\begin{figure}[t]
\centering\includegraphics[width=1.0\linewidth]{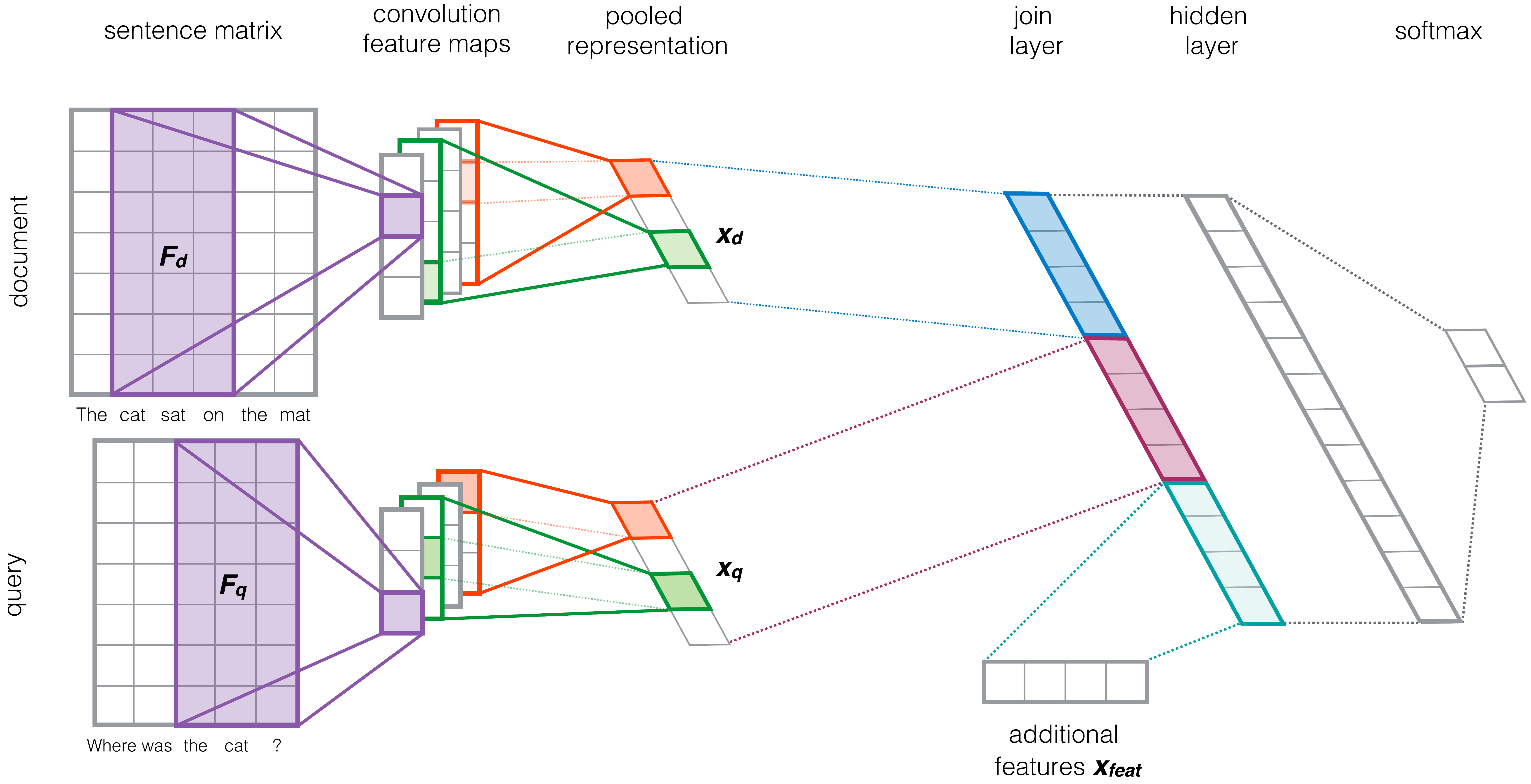}
\caption{The overall design of our convolutional neural network
  for answer selection.}
\label{figure:nn-architecture}
\end{figure}

In this work, we explore the use of convolutional neural networks (CNNs)
for answer selection in an end-to-end question answering task. Our
network is shown in Figure~\ref{figure:nn-architecture}, which is a
slightly simplified version of the model proposed by Severyn and
Moschitti~\cite{Severyn_Moschitti_SIGIR2015}.

We chose to work with this particular CNN for several reasons. It is a simple
model that delivers reproducible results with multiple
implementations~\cite{Rao_etal_SIGIR2017}. It is quick to train (even on CPUs), which supports fast
experimental iteration. Although its effectiveness in answer section is no longer
the state of the art, the model still provides a reasonable baseline (as we show later).

Our model adopts a general ``Siamese'' structure~\cite{bromley-93}
with two subnetworks processing the question and candidate answer (i.e., the ``document'') in
parallel.  This general architecture is fairly common and used in a
variety of other models as
well~\cite{He2016word,he2015multi,he2016umd}. The input to each
``arm'' in the neural network is a sequence of words $[w_1, w_2,
  ... w_{|S|}]$, each of which is translated into its corresponding
distributional vector (i.e., from a word embedding), yielding a
sentence matrix. Convolutional feature maps are applied to this sentence
matrix, followed by ReLU activation and simple max-pooling, to arrive
at a representation vector $\textrm{x}_q$ for the query and
$\textrm{x}_d$ for the candidate answer (``document'').

At the join layer (see Figure~\ref{figure:nn-architecture}), all
intermediate representations are concatenated into a single vector:
\begin{equation}
\textrm{x}_{\textrm{join}} = [ \textrm{x}_q^T; \textrm{x}_d^T; \textrm{x}_{\textrm{feat}}^T ]
\end{equation}

\noindent The final component of the input vector at the join layer consists of
``extra features'' $\textrm{x}_{\textrm{feat}}$ derived from four word
overlap measures between the question and the candidate
sentence:\ word overlap and {\it idf}-weighted word overlap between
all words and only non-stopwords.
As this model is fairly well known,
we refer interested readers to the papers cited above for more details.

\subsection{QA Dataset}

Experiments in this paper use the popular TrecQA dataset for evaluating answer
selection. The TrecQA dataset was first introduced by Wang et
al.~\cite{wang2007jeopardy} and further elaborated by Yao et
al.~\cite{yao2013answer}. The dataset contains a set of factoid
questions, each of which is associated with a number of candidate
sentences that either contain or do not contain the answer (i.e.,
positive and negative examples). The questions come from the Question
Answering Tracks from TREC 8--13~\cite{Voorhees_Tice_TREC8,Voorhees_Dang_TREC2005}, and the candidate answers are
derived from the output of track participants, ultimately drawn from
the collections listed in Table~\ref{collections}. The dataset
comes pre-split into train, development, and test portions, with
statistics shown in Table~\ref{stats}.

\begin{table}[t]
\centering
\resizebox{\columnwidth}{!}{
\begin{tabular}{c|c}
\hline
Dataset                                                             & Document Collections                                                                                                                                                                                                                                               \\ \hline\hline
TREC8                                                            & TREC disks 4\&5 minus Congressional Record                                                                                                                                                                                                             \\ \hline
\begin{tabular}[c]{@{}c@{}}TREC9\\ TREC10\end{tabular}           & \begin{tabular}[c]{@{}c@{}}AP newswire (Disks 1-3)\\ Wall Street Journal (Disks 1-2) \\ San Jose Mercury News (Disk 3)\\ Financial Times (Disk 4) \\ Los Angeles Times (Disk 5) \\  Foreign Broadcast Information Service (FBIS) (Disk 5)\end{tabular} \\ \hline
\begin{tabular}[c]{@{}c@{}}TREC11\\ TREC12\\ TREC13\end{tabular} & AQUAINT disks                                                                                                                                                                                                                                          \\ \hline
\end{tabular}
}
\vspace{0.2cm}
\caption{Source document collections for TrecQA.}
\label{collections}
\end{table}

\begin{table}[t]
\centering
\begin{tabular}{c|r|r|r}
\hline
Set   & \# Question & \# Pos Answers & \# Neg Answers \\ \hline\hline
Train & 1,229       & 6,403         & 47,014        \\ \hline
Dev   & 82         & 222          & 926          \\ \hline
Test  & 100         & 284          & 1,233         \\ \hline \hline
All   & 1,411       & 6,909         & 49,173        \\ \hline
\end{tabular}
\vspace{0.2cm}
\caption{Statistics for various splits of TrecQA.}
\label{stats}
\end{table}

\section{Experiments}

\subsection{Answer Selection Baseline}

We implemented the convolutional neural network shown in
Figure~\ref{figure:nn-architecture} using the PyTorch deep learning
toolkit. Our implementation, which we make available open 
source,\footnote{\url{http://castor.ai/}} is based on a reproducibility study of
Severyn and Moschitti's model~\cite{Severyn_Moschitti_SIGIR2015} by
Rao et al.~\cite{Rao_etal_SIGIR2017} using the Torch deep learning
toolkit (implemented in Lua).\footnote{\url{https://github.com/castorini/SM-CNN-Torch}} 
In fact, our network architecture uses the best setting, as
determined by Rao et al.~via ablation analyses. In particular, they
found that the bilinear similarity component actually decreases effectiveness,
and therefore is not included in our model.

\begin{table}[t]
\centering
\begin{tabular}{lll}
\hline
\textbf{Method} 		& {\bf MAP} & {\bf MRR}\\
\hline\hline
Word overlap				& 0.6496	& 0.6811 \\
{\it idf}-weighted word overlap		& 0.7014	& 0.7688 \\
Our CNN model	    		& 0.7400	& 0.8131 \\
Rao et al.~\cite{Rao_etal_CIKM2016}              & 0.780         & 0.834 \\
\hline
\end{tabular}
\vspace{0.2cm}
\caption{Results comparing our baselines, our CNN model, and the state
  of the art on the TrecQA dataset.}
\label{tab:baselines}
\vspace{-0.3cm}
\end{table}

We adopted the same experimental procedures and settings as Rao et al.~\cite{Rao_etal_SIGIR2017} and
report effectiveness on the TrecQA dataset in
Table~\ref{tab:baselines}. Against this CNN for answer selection, we
compared two very simple baselines:

\begin{itemize}[leftmargin=*]

\item \textbf{Word overlap}, which is the count of how many words in
  the question also appear in the answer candidate (after removing
  stopwords).

\item \textbf{{\it idf}-weighted word overlap}, which is the same
  measure as above, except that matches are weighted with the {\it
    idf} value of the question word.

\end{itemize}

\noindent The main takeaway from these results is that our CNN is only
about 6\% more effective than a simple {\it idf}-based matching
technique. In other words, our convolutional neural network is ``doing
a lot'' for not much gain.

Let's take a step back and consider the broader context of these
results. We can consult an ACL wiki page that nicely summarizes the
state of the art in this answer selection task on the TrecQA dataset~\cite{ACLwiki}. In
Table~\ref{tab:baselines}, we show the best reported results as of this
writing, which are the figures published by Rao et
al.~\cite{Rao_etal_CIKM2016} (note this is a different paper than the
one cited above). We make two interesting observations:

\begin{itemize}[leftmargin=*]

\item The state of the art (based on deep learning) is a measly 11\%
  more effective than the simple baseline that uses {\it idf}-weighted
  word overlap.

\item According to the ACL wiki page~\cite{ACLwiki}, which has charted
  the advance of the state of the art over the past decade or so, the
  simple {\it idf}-weighted word overlap approach is better than
  anything reported in the literature until around 2013.

\end{itemize}

\noindent Despite substantial effort (primarily by the natural language
processing community) in applying deep
learning to tackle answer selection, the gains are modest at best on
this dataset. This is somewhat disappointing given the promise of deep
learning, and the gains that we observe are far less impressive than
improvements reported for computer vision tasks and speech recognition.

We are not the first to observe the high effectiveness of the {\it
  idf}-weighted word overlap baseline~\cite{Yih_etal_ACL2013}, although this finding
is not as well known in the community and well reported in the literature as it should
be. Comparison against appropriate baselines is an important component
of evaluation design to ensure that reported gains are not
illusory~\cite{Armstrong_etal_CIKM2009}.

\subsection{End-to-End Evaluation}

Typically, in a pipeline architecture, component-level improvements in
effectiveness may not translate into end-to-end effectiveness
improvements due to the effects of compounding errors and the fact
that bottlenecks lie elsewhere. Given the answer selection results
reported above, we wondered how our convolutional neural network
would fare in an end-to-end evaluation.

For these experiments, we implemented a
multi-stage architecture similar to the one shown in Figure~\ref{fig:e2e-pipeline}. To start, we
used our Anserini retrieval toolkit~\cite{Yang_etal_SIGIR2017},\footnote{\url{http://anserini.io/}} which is built on the
open-source Lucene search engine, to index the collections in
Table~\ref{collections}. Each question was used as a bag-of-words
query to retrieve the top $h$ hits using BM25. All documents were
then segmented into sentences, and we compared the two following conditions:

\begin{itemize}[leftmargin=*]

\item {\it idf}-reranking. All retrieved sentences are reranked using
  {\it idf}-weighted word overlap. The top $k$ are considered for
  evaluation.

\item {\it idf}+CNN-reranking. All retrieved sentences are first
  reranked using {\it idf}-weighted word overlap. The top $k$ are then
  reranked by our CNN answer selection model. All $k$ resulting
  reranked sentences are considered in the evaluation.

\end{itemize}

\noindent There are two wrinkles in our experimental setup. First, although
the TrecQA dataset was ultimately constructed from TREC evaluations,
the provenance information connecting answer candidates to their
source documents does not exist. That is, we do not actually know
which sentences from the original collection are relevant or not
relevant. Of course, we do have the annotated sentences from the
TrecQA dataset, but due to tokenization and other
sentence processing differences, an exact string match is not
sufficient. For example, a candidate answer from the TrecQA dataset appears as:

\begin{quote}
In 1820 , the founder of 
modern nursing , Florence	Nightingale , was born in Florence , Italy .
\end{quote}

\noindent The actual source sentence from the collection is
as follows:

\begin{quote}
On this date: In 1820, the founder of modern nursing, Florence Nightingale, was born in Florence, Italy.
\end{quote}

\noindent We address this issue by computing the Jaccard similarity
between retrieved sentences from the collection and sentences in
the TrecQA dataset for which we have relevance judgments. If we find a matching
sentence with Jaccard similarity above 0.7, we use the judgment of the
matching sentence from the TrecQA dataset. If there is more than one
match, we take the judgment with the highest score.

This simple matching technique enables end-to-end QA evaluation based
on the TrecQA judgments, but highlights the second major issue with
our evaluation:\ missing judgments. Document retrieval followed by
reranking identifies many sentences for which we have no relevance
judgments. These results are shown in Table~\ref{tab:map-mrr-rbp:idf}
for {\it idf}-reranking and Table~\ref{tab:map-mrr-rbp:sm}
for {\it idf}+CNN-reranking. In both cases, we evaluate on the top 200 ranked
documents ($h=200$) from the collection, reporting MAP, MRR, and rank-biased precision
(RBP)~\cite{moffat2008rank} with residuals in parentheses for different values of $k$.
The final column in both tables shows the number of unjudged documents in the
test set (which contains 100 questions).

Due to the sparsity of judgments, the absolute scores are low, and
furthermore it is not clear if our CNN is actually more effective than
{\it idf}-weighted word overlap! At least from these numbers, the
gains from the CNN model in a component-level evaluation
(Table~\ref{tab:baselines}) seemed to have disappeared.

\begin{figure}[t]
\centering\includegraphics[width=1.0\linewidth]{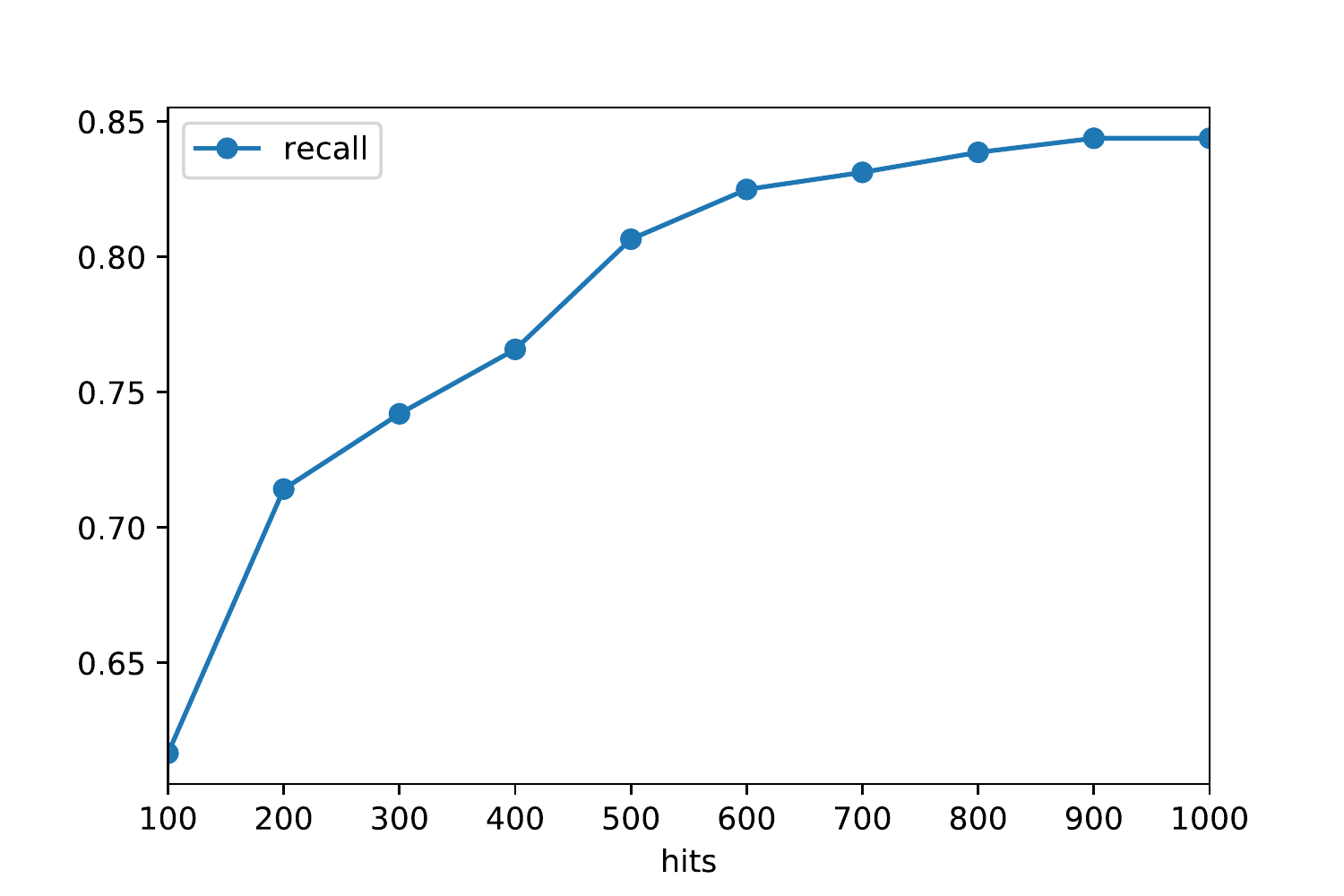}
\caption{Recall of relevant sentences from TrecQA with different
  numbers of documents retrieved.}
\label{figure:sm_recall}
\end{figure}

As a sanity check, sentence-level recall (with respect to the relevant
sentences in the TrecQA dataset) is shown in
Figure~\ref{figure:sm_recall} for different values of $h$ (number of
hits retrieved). The document retrieval component is indeed identifying
relevant candidates, but so many unjudged sentences are brought
into high ranks by the subsequent reranking components that we are
unable to discriminate end-to-end effectiveness using standard
retrieval metrics.

Let us design an evaluation setup that has the best chance of
discriminating between the effectiveness of the CNN and our baseline.
For this, we turn to b-pref~\cite{buckley2004retrieval}, which was
specifically created to handle cases with missing
judgments. Furthermore, instead of evaluating only the top $k$
results, we consider all sentences returned. That is, we rank and
evaluate {\it all} sentences in the top $h$ hits---once again,
comparing {\it idf}-reranking and {\it idf}+CNN-reranking. This setup
maximizes the opportunity for pairwise comparisons that b-pref depends
on.

\begin{figure}[t]
\centering\includegraphics[width=1.0\linewidth]{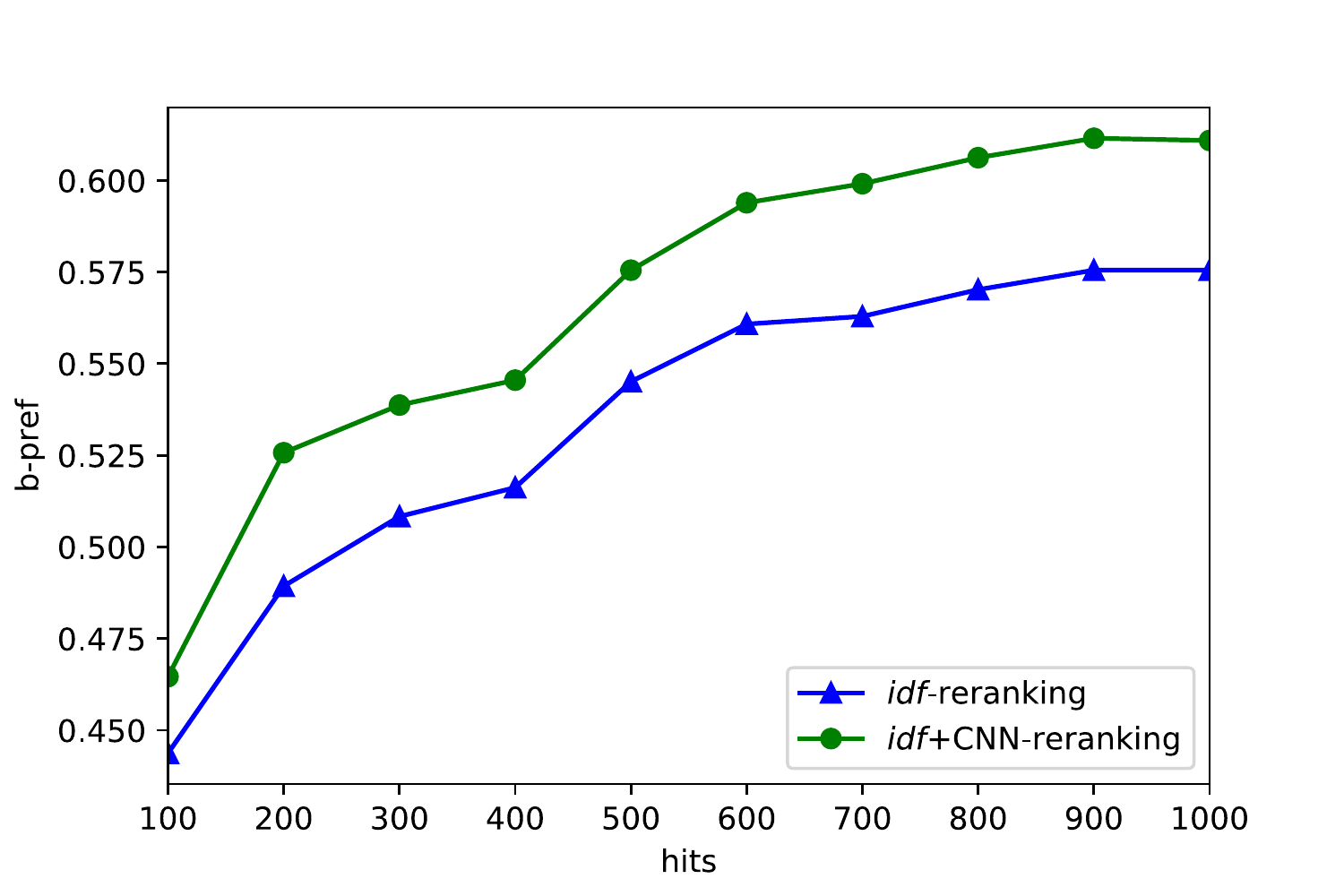}
\caption{B-pref comparisons between {\it idf}-reranking and {\it
    idf}+CNN-reranking with different numbers of documents retrieved.}
\label{figure:sm_bpref}
\end{figure}

The results of this experiment are shown in
Figure~\ref{figure:sm_bpref}. Here, we see indeed that the CNN
effectiveness appears to beat the baseline, but this doesn't
capture the user's perspective when interacting with a
QA system. We are able to obtain discrimination between the two
techniques only by reranking a large number of candidate
sentences---in reality, however, users only care about the top few results in a
QA system's output. In a more reasonable setup of $k=5$ and $h=200$,
{\it idf}-reranking produces a b-pref score of 0.1590 and {\it
  idf}+CNN-reranking produces a b-pref score of 0.1593, which are for
all practical intents indistinguishable.

\begin{table}[t]
\centering
\begin{tabular}{lrrrr}
\hline
\textbf{k} & {\bf MAP} & {\bf MRR} & {\bf RBP (p = 0.5)} & {\bf unjudged}\\
\hline\hline
5000	& 0.1261	& 0.1901	& 0.0903 (0.8735)  & 478612\\
1000	& 0.1259	& 0.1900	& 0.0903 (0.8735)  & 99061\\
500	    & 0.1259	& 0.1900	& 0.0903 (0.8735)  & 49305\\
100	    & 0.1245	& 0.1893	& 0.0903 (0.8735)  & 9675\\
50	    & 0.1216	& 0.1883	& 0.0903 (0.8735)  & 4785\\
10	    & 0.1045	& 0.1775	& 0.0902 (0.8736)  & 914\\
\hline
\end{tabular}
\vspace{0.2cm}
\caption{MAP, MRR, and RBP (residuals in parentheses) for end-to-end
  QA using {\it idf}-reranking (with the number of documents retrieved $h=200$).}
\label{tab:map-mrr-rbp:idf}
\end{table}

\begin{table}[t]
\centering
\begin{tabular}{lrrrr}
\hline
\textbf{k} & {\bf MAP} & {\bf MRR} & {\bf RBP (p = 0.5)} & {\bf unjudged} \\
\hline\hline
5000  & 0.1011	& 0.1721	& 0.0870 (0.8985) & 477505\\
1000  &	0.1146	& 0.2029	& 0.1066 (0.8763)  & 99032\\
500	  & 0.1161	& 0.2035	& 0.1066 (0.8775)  & 49289\\
100	  & 0.1208	& 0.2102	& 0.1103 (0.8709)  & 9674\\
50	  & 0.1260	& 0.2228	& 0.1192 (0.8558)  & 4785\\
10	  & 0.1138	& 0.2314	& 0.1238 (0.8303)  & 916\\
\hline
\end{tabular}
\vspace{0.2cm}
\caption{MAP, MRR, and RBP (residuals in parentheses) for end-to-end
  QA using {\it idf}+CNN-reranking (with the number of documents
  retrieved, $h=200$).}
\label{tab:map-mrr-rbp:sm}
\end{table}

\subsection{Manual Assessment}

Summarizing the results so far:\ it is not clear if our convolutional
neural network is actually more effective than the {\it idf}-weighted
word overlap baseline according to standard retrieval metrics. Given
that the differences in effectiveness are already modest in the answer
selection task, it is entirely possible that the differences are
``swamped out'' by the document retrieval component.

To further examine this issue, we performed a manual assessment of the
answers returned by both the {\it idf}-reranking and the {\it
  idf}+CNN-reranking conditions. We adopted a fairly standard setup
(cf.~\cite{Sanderson_etal_SIGIR2010,Wang_etal_SIGIR2015}) where the
top $k$ results from both conditions are shown to a human assessor in
a side-by-side format. Which side (left or right) displayed which
condition was randomized and blinded from the assessor to ensure an
unbiased evaluation. For each question, the assessor could select from
four judgments:

\begin{itemize}[leftmargin=*]

\item {\it Left.} The assessor prefers the answers on the left.

\item {\it Right.} The assessor prefers the answers on the right.

\item {\it Both.} The assessor expresses no preference; both answers are equally good.

\item {\it Neither.} The assessor expresses no preference; both answers are equally bad.

\end{itemize}

\noindent In this manual evaluation, we arbitrarily set $h$ (number of documents retrieved) to 200 and
evaluated the top five ($k=5$) answers. Manual assessment results by
two of the co-authors are shown in Table~\ref{tab:judging}.

Based on the Wilcoxon sign test (which takes into account ties) as
well as the binomial test (where ties are discarded), we find that
{\it idf}+CNN-reranking is more effective than {\it idf}-reranking
($p<0.05$). In other words, deep learning is contributing to a 
human-detectable improvement in question answering effectiveness.

Interestingly, we find that inter-annotator agreement between the two
assessors is only 0.4103 in terms of Cohen's $\kappa$, which can be
characterized as moderate. This means that although the assessors agree
that {\it idf}+CNN-reranking is more effective than {\it
  idf}-reranking, they don't necessarily agree on {\it which answers}
are better.

\begin{table}[t]
\centering
\begin{tabular}{lrr}
\hline
\textbf{Configuration} 			& {\bf Judge1} & {\bf Judge2}\\
\hline\hline
{\it idf}+CNN-reranking 		& 30		   & 39\\
{\it idf}-reranking  			& 17           & 18\\
Both	    					& 14 		   & 11\\
Neither	    					& 39 		   & 32\\
\hline
\end{tabular}
\vspace{0.2cm}
\caption{Manual assessment of the end-to-end QA results, considering
  the top $k=5$ answers (with the number of documents retrieved $h=200$)}
\label{tab:judging}
\end{table}

\section{Conclusions}

The ultimate goal of a question answering system is to address a
user's information need, and thus it is important to evaluate a system
from an end-to-end perspective. The literature, however, has almost
exclusively focused on the answer selection task, which is only one
component in a standard QA pipeline. Even evaluated in isolation, the
gains that have been achieved by deep learning techniques are modest
at best. However, a manual evaluation appears to show that these
gains {\it do} translate into human-detectable improvements in end-to-end answer
quality.

\section{Acknowledgments} 

This work was supported by the Natural Sciences and Engineering
Research Council (NSERC) of Canada, with additional contributions from
the U.S.\ National Science Foundation under CNS-1405688. Any findings,
conclusions, or recommendations express\-ed do not necessarily reflect
the views of the sponsors.

%\balance


\end{document}